# Impact of Nonuniform Thermionic Emission on the Transition Behavior between Temperature- and Space-Charge-Limited Emission

Dongzheng Chen[1], Ryan Jacobs[1], Dane Morgan[1], and John Booske[2,*]

[1]Department of Materials Science and Engineering, University of Wisconsin-Madison, Madison, WI 53706, USA

[2]Electrical and Computer Engineering Department, University of Wisconsin–Madison, Madison, WI 53706, USA

*Corresponding author. E-mail: jhbooske@wisc.edu

**Abstract:**

Experimental observations have long-established that there exists a smooth *roll-off* or *knee* transition between the temperature-limited (TL) and full-space-charge-limited (FSCL) emission regions of the emission current density-temperature $J - T$ (Miram) curve, or the emission current density-voltage $J - V$ curve for a thermionic emission cathode. In this paper, we demonstrate that this experimentally observed smooth transition does not require frequently used *a priori* assumptions of a continuous distribution of work functions on the cathode surface. Instead, we find the smooth transition arises as a natural consequence of the physics of nonuniform thermionic emission from a spatially heterogeneous cathode surface. We obtain this smooth transition for both $J - T$ and $J - V$ curves using a predictive nonuniform thermionic emission model that includes 3-D space charge, patch fields (electrostatic potential nonuniformity on the cathode surface based on local work function values), and Schottky barrier lowering physics and illustrate that a smooth knee can arise from a thermionic cathode surface with as few as two discrete work function values. Importantly, we find that the inclusion of patch field effects is crucial for obtaining accurate $J - T$ and $J - V$ curves, and the further inclusion of Schottky barrier lowering is needed for accurate $J - V$ curves. This finding, and the emission model provided in this paper have important implications for modeling electron emission from realistic, heterogeneous surfaces. Such modeling is important for improved understanding of the interplay of emission physics, cathode materials engineering, and design of numerous devices employing electron emission cathodes.

**Introduction:**

Thermionic cathodes provide the electron source in numerous vacuum electronic devices (VEDs) applied to civilian, industrial, and scientific applications, such as communication devices, electron microscopes, electron beam lithography, ion thrusters, thermionic energy converters, and free electron lasers.[1], [2] The theory of thermionic emission has been studied for more than a century. The simplest physical model for a thermionic cathode is a perfectly smooth cathode surface with a single work function value, referred to as a "uniform cathode" in this article. The physics of the thermionic emission from an infinitely large uniform cathode in a parallel diode has been thoroughly studied. The Richardson-Laue-Dushman equation[3], [4] with Schottky barrier lowering[5] describes the temperature-limited (TL) emission current density of a uniform cathode. The Child-Langmuir law[6], [7] and Langmuir and Fry's studies[7], [8] provide a model of the full-space-charge-limited (FSCL) emission. Scott's and Eng's works[9], [10] unified both the effects of Schottky barrier lowering and space charge and therefore are able to predict not only the TL and FSCL regions but also the TL-FSCL transition region for a uniform cathode. However, such predicted TL-FSCL transition from a uniform cathode is sharp, qualitatively different from experimental results of actual thermionic cathode emission measurements,[10] which are characterized by a smooth, more gradual TL-FSCL transition in the Miram and the $I - V$ curves.

Experimental results including thermionic electron emission microscopy (ThEEM) images reveal that polycrystalline cathodes have a spatial distribution of work function and emit nonuniformly.[11]–[14] It is known that the nonuniform thermionic emission in a parallel diode is subject to the effects of 3-D space charge[15]–[17], patch fields[11], [12], Schottky barrier lowering[5], and the lateral motion of electrons[15], [18], [19]. By patch field effect, we refer to a nonuniform electrostatic potential on the cathode surface based on local work function values.[11], [12], [20]–[22] This local, nonuniform, electrostatic potential distribution results from the microscopic, lateral transfer of conduction electrons between grains having different work functions, in order to equalize the Fermi energy between all contacting, conducting grains. Each of these effects has been studied in detail separately. While there is still no general, physics-based model unifying all of these effects, previous efforts have made advances in combining some subsets of the effects together. For example, the theory of the anomalous Schottky effect unifies the effects of patch fields and Schottky barrier lowering,[23] and recent studies[15], [19] discuss the effects of 3-D space charge and the lateral motion of electrons. A key result from the work of Chernin *et al.*[15] is that the lateral motion of electrons has only a minor effect on predictions of averaged emission current density from an infinitely large cathode in an infinite parallel diode when the 3-D space charge effect is considered. Therefore, it is possible to accurately predict the averaged emission current density under the assumption that the electrons are restricted to move one-dimensionally from the cathode to the anode with no lateral momentum, which is equivalent to assuming an infinite magnetic field is present along the anode-cathode direction.

In this article, we develop a physics-based nonuniform thermionic emission model unifying the effects of 3-D space charge, patch field effects, and Schottky barrier lowering. Based on the work from Chernin *et al.*[15], we neglect the lateral motion of electrons. We note here that the lateral motion of electrons may have a larger effect in some extreme cases, for example where the emitting patches of the cathode have a length scale comparable to the anode-cathode distance, but we do not consider that case here. Our present model is able to predict the entire domain of the $J - T$ (Miram) and the $J - V$ curves including the TL-FSCL transition for a heterogeneous cathode surface in a perfect infinite parallel diode. Our model results demonstrate that the smooth and gradual TL-FSCL transitions observed in experiments can be reproduced by the model and are natural consequences of the physics of nonuniform thermionic emission. We are therefore able to reproduce the experimentally observed smooth and gradual TL-FSCL transition without relying on any empirical equations such as the Longo-Vaughan equation[24], [25] or an *a priori* assumption of a continuous distribution of work functions[26] on the cathode surface.

A physics-based emission model unifying all important effects has important implications for studying nonuniform thermionic emission as it enables the in-depth understanding of the impact of nonuniform thermionic emission on the TL-FSCL transition. This type of modeling can increase understanding of the interplay of different contributions to emission physics, the materials engineering of cathodes including not only bulk polycrystalline cathodes but also novel cathodes like those based on 2D materials (e.g. graphene) [17], [27], and the design of devices employing electron emission cathodes.

**Methods:**

For the model of thermionic emission from heterogeneous surfaces developed in this work, the cathode is located at $z = 0$, and the anode at $z = d$, where $d$ is the anode-cathode distance. It is assumed that the distribution of the electron energies follows a Maxwell-Boltzmann distribution, and that the motion of the electrons between the cathode and the anode follows non-relativistic classical electrodynamic behavior. The total energy for each electron is conserved: $E = E_{\mathrm{p}}(x, y, z) + mv^2/2$, where $E_{\mathrm{p}}(x, y, z)$ is the potential energy at position $(x, y, z)$, $m$ is the mass of an electron, and $v$ is the velocity. In the absence of an energy barrier at the surface of the cathode, the emission current density at position $(x, y)$ due to electrons of energy between $E$ and $E + \mathrm{d}E$ has the form:[10]

$$\mathrm{d}J(x, y; E) = \frac{AT}{k} \exp\left(-\frac{E - E_{\mathrm{F}}(x, y)}{kT}\right) \mathrm{d}E \qquad (1)$$

where $A = 4\pi m e k^2/h^3$ is the Richardson constant, $e$ is elementary charge, $k$ is Boltzmann's constant, $h$ is Planck's constant, $T$ is temperature, and $E_{\mathrm{F}}(x, y)$ is the local Fermi energy level of the cathode.

The model for the steady state electron density depends on the shape of the potential energy that is created under steady state emission. Figure 1 shows sketches of the potential energy $E_{\mathrm{p}}(x, y, z')$ for a given $(x, y)$ as a function of $z'$ for different cases. For a given position $(x, y, z)$, we define the cathode-side barrier as $E_-(x, y, z) = \max_{0 \le z' \le z} E_{\mathrm{p}}(x, y, z')$ and the anode-side barrier as $E_+(x, y, z) = \max_{z \le z' \le d} E_{\mathrm{p}}(x, y, z')$. The values of $E_-$ and $E_+$ determine how many electrons are able to reach the position $(x, y, z)$, in the case that the lateral motion of electrons and quantum effects are neglected.

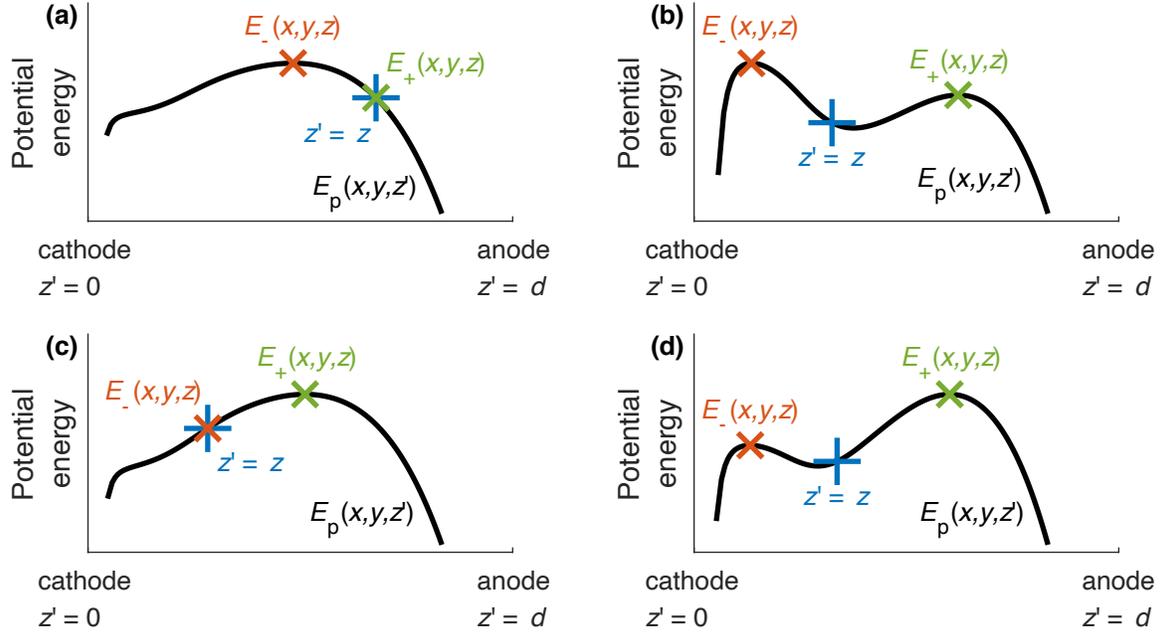

**Figure 1.** Sketch of the potential energy $E_p(x, y, z')$ for a given $(x, y)$ as a function of $z'$ (black curve). For a given position $(x, y, z)$ (blue plus mark), its cathode-side barrier $E_-(x, y, z)$ is marked as red cross, and its anode-side barrier $E_+(x, y, z)$ as green cross. (a) and (b) are examples for the case of $E_-(x, y, z) \geq E_+(x, y, z)$, while (c) and (d) for $E_-(x, y, z) < E_+(x, y, z)$.

Considering the positions $(x, y, z)$ satisfying $E_-(x, y, z) \geq E_+(x, y, z)$, electrons emitted from the cathode surface at position $(x, y, 0)$ with energy $E \geq E_-$ can pass through the cathode-side barrier and reach the position $(x, y, z)$ as they move toward the anode. The charge density for positions $(x, y, z)$ where $E_-(x, y, z) \geq E_+(x, y, z)$ has the form:

$$\rho(x, y, z) = -\int_{E=E_-}^{\infty} \frac{\mathrm{d}J}{v} = -\int_{E=E_-}^{\infty} \frac{\mathrm{d}J}{\sqrt{2(E - E_p)/m}}, \quad E_-(x, y, z) \geq E_+(x, y, z) \tag{2}$$

where the electron velocity $v = \sqrt{2(E - E_p)/m}$.

However, for any position $(x, y, z)$ where $E_-(x, y, z) < E_+(x, y, z)$, electrons emitted from the cathode surface at $(x, y, 0)$ with energy $E \geq E_-$ can still pass through the cathode-side barrier and reach the position $(x, y, z)$. Electrons with energy $E_- \leq E < E_+$ will be reflected back toward the cathode by the higher anode-side barrier and pass through the location $(x, y, z)$, this time moving back toward the cathode. Therefore, the charge density for positions $(x, y, z)$ where $E_-(x, y, z) < E_+(x, y, z)$ has the form:

$$\rho(x,y,z) = -\left[\int_{E=E_-}^{\infty} \frac{\mathrm{d}J}{\sqrt{2(E-E_\mathrm{p})/m}} + \int_{E=E_-}^{E_+} \frac{\mathrm{d}J}{\sqrt{2(E-E_\mathrm{p})/m}}\right], \quad E_-(x,y,z) < E_+(x,y,z) \tag{3}$$

Substituting Equation 1 into Equations 2 and 3, we obtain a closed-form expression of the relation between the potential energy $E_\mathrm{p}(x,y,z)$ and the charge density $\rho(x,y,z)$:

$$\rho(x,y,z) =$$
$$\begin{cases} -\sqrt{\frac{\pi m}{2kT}} AT^2 \exp\left(-\frac{E_\mathrm{p}(x,y,z)-E_\mathrm{F}(x,y)}{kT}\right) \mathrm{erfc}\left(\sqrt{\frac{E_-(x,y,z)-E_\mathrm{p}(x,y,z)}{kT}}\right), \quad E_-(x,y,z) \geq E_+(x,y,z) \\ -\sqrt{\frac{\pi m}{2kT}} AT^2 \exp\left(-\frac{E_\mathrm{p}(x,y,z)-E_\mathrm{F}(x,y)}{kT}\right)\left[2\,\mathrm{erfc}\left(\sqrt{\frac{E_-(x,y,z)-E_\mathrm{p}(x,y,z)}{kT}}\right) - \mathrm{erfc}\left(\sqrt{\frac{E_+(x,y,z)-E_\mathrm{p}(x,y,z)}{kT}}\right)\right], \quad E_-(x,y,z) < E_+(x,y,z) \end{cases} \tag{4}$$

where erfc is the complementary error function.

The electrostatic potential $V$ and the charge density $\rho$ satisfy Poisson's equation, and therefore the 3-D space charge effect is included in this model through:

$$\nabla^2 V(x,y,z) = -\frac{\rho(x,y,z)}{\epsilon_0} \tag{5}$$

where $\epsilon_0$ is the vacuum permittivity.

The boundary condition for the cathode surface is[10], [21]–[23]

$$V(x,y,z=0) = -\frac{E_\mathrm{F}(x,y)+\phi(x,y)}{e} \tag{6}$$

where $E_\mathrm{F}(x,y)$ and $\phi(x,y)$ are the local Fermi level and the local work function of the cathode, respectively.

Similarly, the boundary condition for the anode surface is[10]

$$V(x,y,z=d) = -\frac{E_\mathrm{F,anode}(x,y)+\phi_\mathrm{anode}(x,y)}{e} \tag{7}$$

where $E_\mathrm{F,anode}(x,y)$ and $\phi_\mathrm{anode}(x,y)$ are the local Fermi level and the local work function of the anode, respectively.

At thermodynamic equilibrium, the Fermi level is equal throughout a conductor. In the case of a conductive cathode and a conductive anode, $E_\mathrm{F}(x,y)$ is a constant value throughout the cathode and $E_\mathrm{F,anode}(x,y)$ is a constant value throughout the anode. Neglecting the thermoelectric effect, which is no larger than tens of millivolts under typical operating conditions and negligible compared to the typical applied anode-cathode voltage, the anode-cathode Fermi level difference is $E_{\text{F,anode}} - E_{\text{F}} = -eV_{\text{AK}}$, where $V_{\text{AK}}$ is the anode-cathode voltage as measured by a voltmeter in experiments. For the present model, if we let $E_{\text{F}} = 0$ then $E_{\text{F,anode}} = -eV_{\text{AK}}$. In this case, the boundary conditions are $V(x, y, z = 0) = -\phi(x, y)/e$ and $V(x, y, z = d) = V_{\text{AK}} - \phi_{\text{anode}}(x, y)/e$. The cathode surface is a non-equipotential surface, and the patch field effect is included.

Considering the Schottky effect near the cathode surface, the potential energy $E_p$ takes the form:

$$E_p(x, y, z) = -eV(x, y, z) - \frac{e^2}{16\pi\epsilon_0 z} \tag{8}$$

where $-e^2/(16\pi\epsilon_0 z)$ is the energy term representing the image charge effect.

The nonuniform thermionic emission model can be solved by solving the system of Equations 4-8. A numerical method to solve this system of nonlinear equations is to solve them by iterations of $E_p \rightarrow \rho \rightarrow V \rightarrow E_p$ until convergence is obtained. One of the algorithms to solve Poisson's equation in the step $\rho \rightarrow V$ involves a Fourier transformation for the $x$ and $y$ directions and the Thomas algorithm for the $z$ direction.[15], [16] Once $E_p(x, y, z)$ is solved, one is able to calculate the maximum barrier $E_{\text{p,max}}(x, y) = \max\limits_{0 \leq z \leq d} E_p(x, y, z)$, and the corresponding local emission current density is

$$J(x, y) = AT^2 \exp\left[-\frac{E_{\text{p,max}}(x, y) - E_{\text{F}}(x, y)}{kT}\right] \tag{9}$$

The averaged emission current density of the cathode can be obtained by averaging $J(x, y)$ over the whole cathode surface.

In the following results, we analyze the predicted emission from our model and evaluate how each physical effect impacts the emission by comparing the results for the cases where some subset or all of the three physical effects are considered. We note here that all three effects of 3-D space charge, patch fields, and Schottky barrier lowering are physically present in nonuniform thermionic emission. Although neglecting any of them will make the emission model less physical, we show the results where some of the effects are neglected to illustrate the impact of each effect on the resulting emission.

The patch field effect was neglected in previous studies which included the 3-D space charge effect. [15], [16], [19] To evaluate the importance of the patch field effect, we show the results under a no-patch-field situation equivalent to the methods used in those previous studies [15], [16], [19]. The patch field effect originates from the fact that patches with different work functions will have different local (near-surface) vacuum level energies (different electrostatic potentials on the cathode surface). When placed in electrically conducting contact, there will be transfer of conduction electrons between the grains so that all Fermi energies are made equal. This electron transfer results in local nonuniform electrostatic potential values in the vacuum, immediately above the grains having initially different work functions. This effect is very much related to the contact potential difference that arises at an interface between metals with dissimilar work functions. Grains starting with an initially higher work function (lower Fermi energy) will acquire slightly greater local electron charge and acquire a local electrostatic potential depression relative to their lower work function neighboring grains, which results in the electrostatic potential nonuniformity on the cathode surface based on local work function values. To correctly predict the spatially nonuniform emission energy barrier, one needs to apply a nonuniform electrostatic potential distribution to the surface, where the local "cathode" potential is more or less negative in proportion to the local work function of each grain, relative to the common Fermi level (nominal cathode voltage) of the cathode (Equation 6). Thus, if the nominal cathode voltage is assumed to be at a reference zero volts, then the surface (vacuum) electrostatic potential value of a 2.0 eV work function grain assumes a -2.0 V potential value whereas a 2.5 eV work function grain assumes a -2.5 V potential value. The no-patch-field results are obtained under the assumption that the cathode surface have the same local near-surface vacuum level, i.e., the boundary condition $V(x, y, z = 0)$ is constant throughout the cathode surface, so there will be no patch field effect. In no-patch-field cases, we assume that the boundary condition of the cathode surface is $V(x, y, z = 0) = -\overline{\phi}/e$, where $\overline{\phi}$ is the average value of $\phi(x, y)$ over the whole cathode surface, and that $E_{\mathrm{F}}(x, y) = \overline{\phi} - \phi(x, y)$. Under this assumption, the spatial distribution of work function $-eV(x, y, z = 0) - E_{\mathrm{F}}(x, y) = \phi(x, y)$ is still the same as the with-patch-field cases. The comparison between no-patch-field (common local vacuum-level) and with-patch-field (common Fermi-level) results will illustrate the impact of the patch field effect.

The results without the Schottky effect are obtained by omitting the image charge term $-e^2/(16\pi\epsilon_0 z)$ in Equation 8. The results without space charge are obtained by solving the maximum barrier $E_{p,max}(x, y)$ at absolute zero temperature when there is no space charge, with the assumption that the maximum barrier $E_{p,max}(x, y)$ remains the same at finite temperatures. The 1-D space charge effect cases illustrate the results under the assumption that the space charge forces from different work function patches do not interact with each other. The results considering the 1-D space charge effect without the patch field effect are obtained under the assumption that each patch emits independently, where we replace the 3-D Laplace operator $\nabla^2$ in Equation 5 with the corresponding 1-D operator $\partial^2/\partial z^2$. The results considering the 1-D space charge effect with the patch field effect are obtained in the following order: (1) Solve the model with patch fields included at absolute zero temperature to get the potential energy $E_{p1}$. (2) Solve the model without patch field effects at absolute zero temperature to get the potential $E_{p2}$. (3) Assume the additional potential energy due to patch field is $E_{PF} = E_{p1} - E_{p2}$. (4) Solve the model after replacing the 3-D Laplace operator $\nabla^2$ in Equation 5 with the corresponding 1-D operator $\partial^2/\partial z^2$ and adding a term $E_{PF}$ to the right side of Equation 8.

**Results**

In this work, we use an idealized model heterogeneous surface characterized by an infinitely large, periodic checkerboard spatial distribution of work functions, as shown in Figure 2. Similar checkerboard distributions have been used in many previous studies of nonuniform emission.[11], [12], [16], [19], [28], [29] Here, a checkerboard model surface with work function values of $\phi_1 = 2$ eV and $\phi_2 = 2.5$ eV and with square size $a = 5$ μm is used. These work function values are typical values for sintered porous tungsten (dispenser) cathodes and the square size is the typical grain size.[2], [18], [30], [31] The anode-cathode distance $d$ is chosen to be 1 mm, typical in closely spaced diode tests[32], which is much larger than patch size $a$. The work function of the anode is chosen to be $\phi_{anode}(x, y) = 4.5$ eV, which is close to the work function value of many anode materials including stainless steel, Monel, copper, tungsten, and molybdenum.[33]

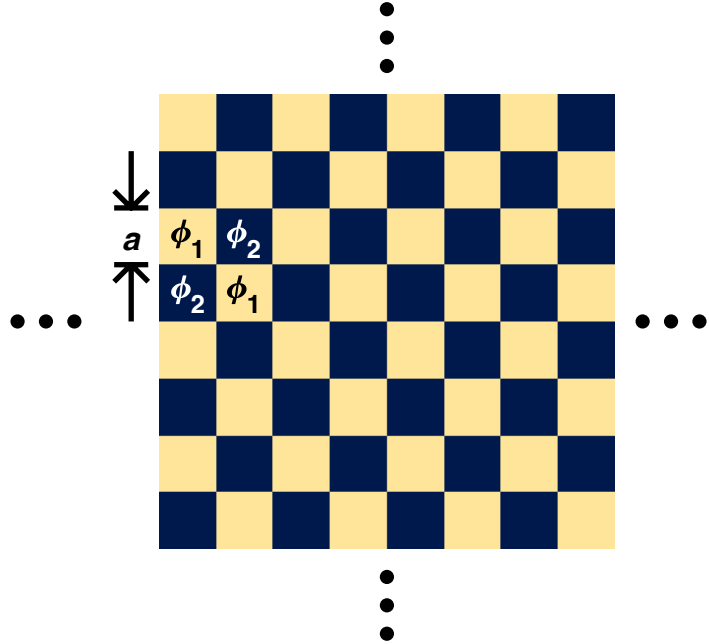

**Figure 2.** Model heterogenous emission surface characterized as a checkerboard of spatial distributions of work function. In this work, values of $\phi_1 = 2$ eV and $\phi_2 = 2.5$ eV and square size $a = 5$ μm are used.

Figure 3 contains predicted $J - T$ (Miram) and $J - V$ (or $I - V$) curves separately showing the effect of space charge at the level of 1-D and 3-D, patch fields, and Schottky barrier lowering on the resulting emission current density. From Figure 3, some qualitative, general features of the current density as a function of $T$ and $V$ emerge based on the inclusion of different physical effects. The inclusion of space charge effects reduces the total emission, where 1-D space charge results in greater reduction in total emission than 3-D space charge. The inclusion of patch fields also reduces the total emission. Finally, the Schottky effect increases the emission by reducing the emission surface barrier. These general findings are consistent with a number of previous studies[5], [7]–[10], [15].

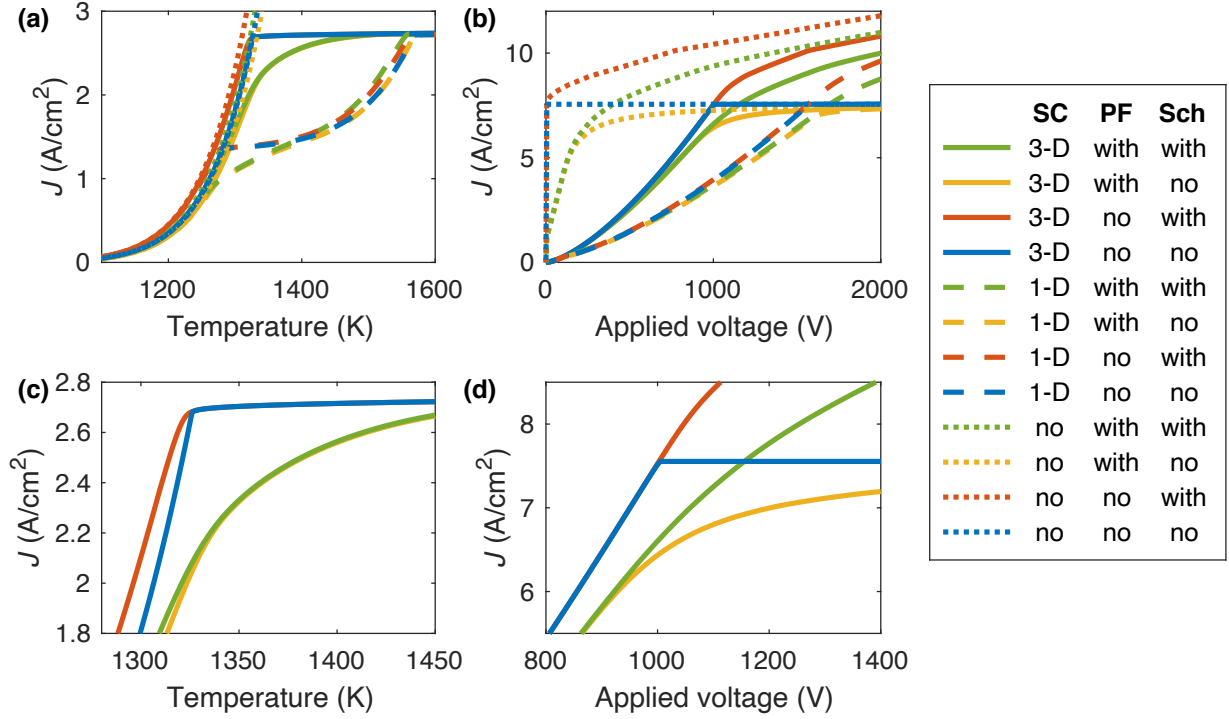

**Figure 3.** Predicted (a) $J - T$ (Miram) curves at applied voltage $V_{AK} = 500$ V and (b) $J - V$ curves at temperature $T = 1400$ K with various combinations of space charge (SC), patch field (PF), and Schottky barrier lowering (Sch) effects considered. (c) and (d) focus on the TL-FSCL transition region of (a) and (b), respectively, for the case of 3-D space charge with and without the effect of patch fields and Schottky barrier lowering.

In the predicted $J - T$ curves (Figure 3a), all of the $J - T$ curves with no space charge effects (dotted curves) increase exponentially. This behavior occurs regardless if the patch field and Schottky effects are considered, and approximately follows the behavior of the Richardson-Laue-Dushman equation. Results of 1-D space charge without the patch field effect (dashed red and dashed blue curves) give peculiarly stepped curves qualitatively inconsistent with experiment, calling into question the assumption that each patch emits independently. Figures 3a and 3c illustrate that the 3-D space charge effect itself (solid blue curve) does not make the TL-FSCL transition region smooth in the condition similar to typical closely spaced diode tests of dispenser tungsten cathodes. However, smooth TL-FSCL transition regions are observed when both 3-D space charge and patch field effects are included together (solid green and solid yellow curves). Compared with space charge and patch field effects, the Schottky effect is minor in determining the shape of the $J - T$ curves, and only acts to make the transition slightly smoother (solid red vs.

solid blue curves, and solid green vs. solid yellow curves, best observed in Figure 3c).

Figure 3b and Figure 3d show the predicted $J - V$ curves. In Figure 3b and Figure 3d, all curves without the Schottky effect (yellow and blue curves) converge to values one would obtain from the Richardson-Laue-Dushman equation at the TL regions (high voltage end). In addition, all curves with the Schottky effect included (green and red curves) show the asymptotic Schottky behavior of the current density at high voltages, as observed in experiments. The curves which ignore space charge effects (dotted curves) give the expected asymptotic behavior at the high voltage limit. Contrary to what was observed for the $J - T$ curves, the green, yellow, and red $J - V$ curves in Figure 3b and Figure 3d show that both patch field and Schottky effects contribute to the smoothness of the TL-FSCL transition. Although the behavior of the $J - T$ curves with 1-D space charge (dashed curves) in Figure 3a differs substantially from typical experimental curves, the behavior of the $J - V$ curves with 1-D space charge effects (dashed curves) in Figure 3b is qualitatively similar with the corresponding $J - V$ curves when 3-D space charge effects are included (solid curves).

Both the $J - T$ and $J - V$ curves are commonly used to evaluate the cathode performance and it is therefore critical for an emission model to accurately predict the behavior of both curves. Among the twelve cases plotted in Figure 3, only the case where all the effects of 3-D space charge, patch fields, and Schottky barrier lowering are considered (solid green curves) predicts a smooth $J - T$ (i.e., a smooth TL-FSCL transition with temperature and a smooth Miram curve knee) and $J - V$ curve with the Schottky behavior, thereby reproducing the known experimental cathode emission behavior as both a function of temperature and applied voltage.

**Conclusion**

Overall, we have shown that for predicting $J - T$ curves (also known as Miram curves), the space charge and the patch field effects (electrostatic potential nonuniformity on the cathode surface based on local work function values) play a more important role than the Schottky effect in determining the shape of the TL-FSCL transition. On the other hand, for predicting $J - V$ curves, the patch field effect and Schottky effect are both essential to predict asymptotic Schottky behavior of the current density at high voltages.

Even with the simple heterogenous work function distribution considered in this work, consisting of only two discrete work function values, the predicted TL-FSCL transition regions are smooth for both the $J - T$ and $J - V$ curves, in agreement with experimental observations on real cathodes. The present model results illustrate that a smooth TL-FSCL transition is a natural consequence of the effects of 3-D space charge, patch fields, and Schottky barrier lowering on nonuniform thermionic emission, and that neither empirical equations such as the Longo-Vaughan equation[24], [25] nor an *a priori* assumption of a continuous distribution of work functions on the emitting surface[26] are necessary to generate a smooth transition. This result suggests that a surface with a set of discrete work functions, perhaps associated with different surface orientation, terminations, and compositions[30], [34], could yield that smooth TL-FSCL behavior seen in actual cathodes.

The insights obtained in this work demonstrate the value of the developed model in predicting the nonuniform thermionic emission from a heterogeneous cathode surface and suggests that the smooth transition behavior observed in experiments may have its origin in the nonuniform emission from relatively simple work function distributions on surfaces. The effects of space charge, patch fields, and Schottky barrier lowering have been studied separately in previous studies. However, this work has unified all of these effects and applied the resulting model to predict the emission current density as a function of temperature and applied voltage for TL and FSCL regions, and, crucially, study the impact of the nonuniformity on the transition region between TL and FSCL regions. We anticipate the emission model presented here will enable further explorations and improved understanding of the interplay of surface physics properties of a cathode and the resulting engineering and design of an array of devices incorporating thermionic cathodes.